\documentstyle[psfig, epsfig, 12pt]{article}
\newfont{\Bbb}{msbm10 scaled 1200}     
\newcommand{\mathbb}[1]{\mbox{\Bbb #1}}

\def\lbldef#1#2{\expandafter\gdef\csname #1\endcsname {#2}}

\def\href#1#2{#2}

\newcommand{\beq}{\begin{equation}}
\newcommand{\eeq}{\end{equation}}
\newcommand{\ber}{\begin{eqnarray}}
\newcommand{\eer}{\end{eqnarray}}
\newcommand{\beqar}{\begin{eqnarray}}

\newcommand{\cA}{{\cal A}}

\newcommand{\cF}{{\cal F}}

\newcommand{\eeqar}{\end{eqnarray}}

\newcommand{\ba}{\begin{eqnarray}}
\newcommand{\ea}{\end{eqnarray}}

\newcommand{\dsl}
  {\kern.06em\hbox{\raise.15ex\hbox{$/$}\kern-.56em\hbox{$\partial$}}}

\newcommand{\eeqarr}{\end{eqnarray}}
\newcommand{\ZZ}{{\rm \kern 0.275em Z \kern -0.92em Z}\;}

\let\a=\alpha \let\b=\beta \let\g=\gamma \let\d=\delta \let\e=\epsilon
    \let\k=\kappa
 \let\m=\mu \let\n=\nu   \let\r=\rho
\let\s=\sigma \let\t=\tau  \let\f=\phi \let\c=\chi 
 \let\G=\Gamma

\def\nn{\nonumber} \def\bd{\begin{document}} \def\ed{\end{document}}
\def\ds{\documentstyle} \let\fr=\frac \let\bl=\bigl \let\br=\bigr
\let\Br=\Bigr \let\Bl=\Bigl 
\let\bm=\bibitem
\let\na=\nabla
\let\pa=\partial \let\ov=\overline 
\newcommand{\be}{\begin{equation}} 
\newcommand{\ee}{\end{equation}} 
\def\ft#1#2{{\textstyle{{\scriptstyle #1}\over {\scriptstyle #2}}}}
\def\fft#1#2{{#1 \over #2}}
\def\vp{\varphi}
\def\sst#1{{\scriptscriptstyle #1}}
\def\oneone{\rlap 1\mkern4mu{\rm l}}
\def\td{\tilde}
\def\wtd{\widetilde}
\def\ie{\rm i.e.\ }
\def\dalemb#1#2{{\vbox{\hrule height .#2pt
        \hbox{\vrule width.#2pt height#1pt \kern#1pt
                \vrule width.#2pt}
        \hrule height.#2pt}}}
\def\square{\mathord{\dalemb{6.8}{7}\hbox{\hskip1pt}}}
\def\wtd{\widetilde}
\def\R{\rlap{\rm I}\mkern3mu{\rm R}}
\def\im{{\rm i}}
\def\tilg{\tilde{g}}
\def\tilF{\tilde{F}}
\def\tilA{\tilde{A}}
\def\varf{\varphi}
\def\tilf{\tilde{\phi}}
\def\tilh{\tilde{h}}
\def\rme{{\rm e}}
\def\cA{{\cal A}}
\def\cF{{\cal F}}
\begin{document}
\baselineskip=15.5pt
\pagestyle{plain}
\setcounter{page}{1}
\begin{titlepage}

\leftline{\tt hep-th/0105019}

\vskip -1.cm

\rightline{\small{\tt CTP-MIT-3129}}
\rightline{\small{\tt CTP-TAMU-12/01}}

\begin{center}

\vskip 0.5 cm

{\LARGE Supergravity duals of gauge field theories from 
$SU(2) \times U(1)$ gauged supergravity in five dimensions}

\vskip 1.cm

{\large Martin Schvellinger$\dagger$ and Tuan A. Tran$^\ddagger$}

\vskip 0.5cm

$^\dagger${\it Center for Theoretical Physics, \\
Laboratory for Nuclear Science and Department of Physics, \\
Massachusetts Institute of Technology, \\
Cambridge, Massachusetts 02139, USA} \\
E-mail: {\tt martin@ctpbeaker.mit.edu}

\vskip 0.5cm

$^\ddagger${\it Center for Theoretical Physics, \\ 
Texas A$\&$M University \\
College Station, Texas 77843, USA} \\
E-mail: {\tt tuan@rainbow.physics.tamu.edu}

\vspace{1cm}

{\bf Abstract}

\end{center}

We study the $SO(4)$-symmetric solution of the five-dimensional $SU(2) \times U(1)$ gauged 
${\cal{N}}=4$ supergravity theory obtained in {\tt [hep-th/0101202]}.
This solution contains purely magnetic non-Abelian and electric Abelian fields. 
It can be interpreted as a reduction of 
seven-dimensional gauged supergravity on a torus, which comes from type IIB
supergravity on $S^3$. We also show how to obtain that solution from
six-dimensional Romans' theory on a circle. We then up-lift the
solution to massless type IIA supergravity. The dual gauge field
theory is twisted and is defined on the worldvolume of a 
NS-fivebrane wrapped on $S^3$. Two other spatial directions of the
NS-fivebrane are on a torus. 
In the IR limit it corresponds to a three-dimensional 
gauge field theory with two supercharges.

\noindent

\end{titlepage}

\newpage


\baselineskip=15.5pt

\section{Introduction}

Finding dual field theories of gauged supergravities with
solutions involving curved manifolds
gives the possibility to explore some aspects of supergravity theories
related to twisted field theories through the well-known AdS/CFT
duality \cite{Maldacena:1998re,Gubser:1998bc,Witten:1998qj}. 
Furthermore, this searching certainly provides new examples of 
the AdS/CFT duality which turn out to be interesting on their own right. 
 
When brane worldvolumes are wrapped on different 
compact spaces 
\cite{Maldacena:2001mw,Maldacena:2001yy,Acharya:2000mu,Nieder:2000kc,Gauntlett:2000ng,
Nunez:2001pt,Edelstein:2001pu}, 
there are several situations where twisted gauge field theories \cite{Bershadsky:1996qy} appear. 
Particularly, fivebranes and
D3-branes wrapped on holomorphic curves were studied 
\cite{Maldacena:2001mw,Maldacena:2001yy}. Also, 
fivebranes \cite{Acharya:2000mu} and D3-branes \cite{Nieder:2000kc} 
wrapped on associative 3-cycles have been investigated, while
extensions to M-fivebranes wrapping K{\"a}hler 4-cycles, special 
Lagrangian 3-, 4- and 5-cycles, co-associative 4-cycles and Cayley 
4-cycles have been systematically
studied in reference \cite{Gauntlett:2000ng}. In a recent paper, 
we have studied supergravity solutions describing the
flows from $AdS_6$-type regions to $AdS_4$ and $AdS_3$ regions, 
by considering the large $N$ limit of D4-branes on 2- and 3-cycles, 
as well as, wrapped NS-fivebranes \cite{Nunez:2001pt}.
 
In this paper, we concentrate on a system which, when up-lifted to 
ten dimensions, can be interpreted as type IIB NS-fivebranes 
wrapped on $S^3 \times T^2$. In particular, this $S^3$ is embedded in
a seven-dimensional $G_2$ holonomy manifold. We consider the decoupling limit of 
$N$ NS-fivebranes wrapped on $S^3 \times T^2$
\cite{Itzhaki:1998dd}, keeping the radii fixed. 
Since the brane worldvolume is curved, 
in order to define covariantly constant Killing spinors, 
the resulting field theory on the brane worldvolume will be twisted. 
In order to describe the flows between the 5+1-dimensional
field theory (defined in the NS-fivebrane worldvolume in the UV) 
and the 2+1-dimensional field theory in the IR, we will start with 
the $SO(4)$-symmetric solution, obtained by \cite{Chamseddine:2001hk} of the 
five-dimensional $SU(2)$ gauged ${\cal{N}}=4$ supergravity
constructed by Romans \cite{Romans:1986ps}. 
That solution only contains magnetic non-Abelian and electric Abelian 
fields. We will show how the five-dimensional supergravity 
can be obtained from a reduction of the seven-dimensional gauged
supergravity theory on a torus. This seven-dimensional theory is
obtained by reducing type IIB supergravity on $S^3$. 
The dual twisted field theory is defined on the NS-fivebrane worldvolume 
wrapped on $S^3$, whereas the other two spatial directions are wrapped 
on a torus. In the IR limit it corresponds to a three-dimensional 
twisted gauge field theory on $R^1 \times T^2$,  
with two supercharges. It is worth noting that this theory does not come
from an $AdS_4$-like manifold since its spatial directions does not 
live in the spatial sector of five-dimensional supergravity, but 
on the torus in 
the ten-dimensional theory. In addition, we will see that in the IR the theory is confining.
On the other hand, the five-dimensional theory can be viewed 
as a reduction of six-dimensional gauged supergravity on a circle.
This relates it to the Romans' six-dimensional $F(4)$ gauged supergravity
theory \cite{Romans:1986tw}. Several aspects of this theory as seen
from the gauge field theory point of view, including some dual twisted 
gauge field theories, have been analyzed in reference \cite{Nunez:2001pt}. 
We then up-lift the previously mentioned solution to massless type 
IIA supergravity on $S^1 \times S^3$. 

If we turn off the electric Abelian fields, it is also possible 
to find a solution for the five-dimensional gauged supergravity \cite{Chamseddine:2001hk}, which
indeed is singular. Using the criterion given in  
reference \cite{Maldacena:2001mw}, one can see that the singularity of
that solution is {\it bad}, so that
in the IR this solution does not represent a gauge field
theory. Therefore, one may say that the electric Abelian fields 
{\it remove} the singularity. It would be interesting to know 
whether the non-singular solution with non-vanishing Abelian 2-form 
is related to the rotation of the NS-fivebrane.
If it were the case, it would probably be related to the mechanism
studied in reference \cite{Maldacena:2000dr}, leading to a desingularization
by rotation.  

This paper is organized as follows. In section 2 we review the basic 
formalism and set-up of the five-dimensional Romans' theories 
\cite{Romans:1986ps}.
The five-dimensional ${\cal{N}}=4$ AdS supergroup is $SU(2,2|2)$ whose 
maximal bosonic supergroup is $SU(2,2) \times SU(2) \times U(1)$.
Furthermore, the $SU(2,2)$ group is isomorphic to $SO(4,2)$ AdS group in 
five dimensions.
The field content and the Lagrangian of this theory will be introduced
in the next section, however here we 
briefly discuss some features of the ${\cal{N}}=4$
supergravity given in \cite{Romans:1986ps}. 
In fact, the gauge group $SU(2) \times U(1)$
generically leads to two coupling constants $g_1$ and $g_2$, 
corresponding to $U(1)$ and $SU(2)$, respectively. In the theories 
analyzed in \cite{Romans:1986ps}, four cases are considered depending 
on the values of $g_2$ and $g_1$. In the 
Romans' paper $g_1$ is always assumed to be non-zero 
because in the kinetic term for the self-dual tensor 
it enters as the factor like $1/g_1$. However, it was pointed out that
the limit $g_1\rightarrow 0$ can be taken after some appropriate
re-scalings and dualization \cite{Cowdall:1999rs,Chamseddine:2001hk}. 
Indeed, we will study a solution having only $SU(2)$ gauge symmetry. 
Hereafter, we will assume $g=g_2$ and $g_1=0$, {\it i.e.} the 
$U(1)$ ungauged theory and there are not two 2-index potentials. 
In section 3, we show how to obtain the 
five-dimensional $SU(2) \times U(1)$ gauged ${\cal{N}}=4$ supergravity from 
seven-dimensional supergravity, which is obtained by reducing type IIB
supergravity on $S^3$, on a torus. We also consider the relation to 
the massless six-dimensional supergravity theory via Kaluza-Klein 
reduction on a circle.
The flow between the 5+1-dimensional gauge theory and the 3-dimensional 
${\cal{N}}=1$ SYM theory on a torus is studied in section 4. 
This flow is driven by the $SO(4)$-symmetric solution, which was
obtained in \cite{Chamseddine:2001hk}, of the 
Romans' theory in five dimensions. Discussion will be given in the
last section.


\section{The Romans' theories in 5 dimensions}


In this section we review the five-dimensional $SU(2)\times U(1)$ gauged  
${\cal {N}}=4$ supergravity constructed by Romans~\cite{Romans:1986ps},
whose conventions we follow. The theory consists of a graviton $e_\m^\a$, 
three $SU(2)$ gauge potentials $A_\m^I$, an $U(1)$ gauge potential 
$\cA_\m$, two 2-index tensor gauge fields $B^\a_{\m\n}$ which
transform as a doublet of $U(1)$, a scalar $\phi$, four gravitinos
$\psi_{\m\, i}$ and four gauginos $\c_\m$. We are interested in the
case in which the $U(1)$ coupling constant and two 2-index
tensor gauge fields are zero. The bosonic
Lagrangian of the theory without the two 2-form potentials and
$U(1)$ gauge coupling is 
\ba
e^{-1}\,{\cal L} &=& -\frac{1}{4} \, R +\frac{1}{2} \, 
(\partial^\mu\phi) \, (\partial_\mu\phi) - \frac{1}{4} \, 
\rme^{\frac{4}{\sqrt{6}}\f}\,F^{I \, \mu\nu} \, F^I_{\mu\nu}
- \frac{1}{4}\,\rme^{-\frac{8}{\sqrt{6}}\f}\,\cA_{\m\n}\,\cA^{\m\n}\nn\\
& & +
\fr18\,g^2\,\rme^{-\frac{4}{\sqrt{6}}\f}
- \frac{1}{4} \,e^{-1}\,\varepsilon^{\m\n\r\s\t}\,F^I_{\m\n} \, F^I_{\r\s}\,
\cA_\t\, \,\,\, ,
\label{romans5}
\ea
where $e$ is the determinant of the vielbein, $g$ is the $SU(2)$
coupling constant and $\varepsilon_{\m\n\r\s\t}$ is a Levi-Civita 
tensor density. The Abelian field strength $\cA_{\m\n}$ and non-Abelian 
field strength $F_{\m\n}^I$ are given by
\ba
\cF_{\m\n} &\equiv& \pa_\m \cA_\n - \pa_\n \cA_\m \,\,\,  , \nn\\ 
F_{\m\n}^I &\equiv& \pa_\m A_\n^I - \pa_\n A_\m^I + g \,
\e^{IJK}\,A^J_\m\,A^K_\n \,\,\, ,
\ea
respectively. The supersymmetry transformations for the gauginos and 
gravitinos are 
\ba
\d\c_a &=& \fr1{\sqrt{2}} \,\g^\m \, (\pa_\m\f) \, \epsilon_a + 
\sqrt{3} \,\, T_{ab} \,\, \epsilon^b - \fr1{2\sqrt{6}}\,\g^{\m\n}\,(H_{\m\n\;ab} - 
\sqrt{2}\,h_{\m\n\;ab})\,\epsilon^b\,\,\, ,\label{gautransf}\\
\d\psi_{\m\, a} & = & {\cal D}_\m\,\epsilon_a  + \g_\m \,\, T_{ab} \,\,\epsilon^b
-\frac{1}{6\sqrt{2}} \, (\gamma_\mu^{\,\,\,\, \nu\rho} - 4 \, 
\delta_\mu^{\,\,\,\, \nu} \, \gamma^\rho)
\left(H_{\nu\rho\;ab} + \fr1{\sqrt{2}}\,h_{\nu\rho\;ab}\right)
\epsilon^b \,\,\, ,
\label{gratransf}
\ea
where $T_{ab}$, \, $H_{\m\n\;ab} \,$ and $h_{\m\n\;ab} \,$ are
defined as follows
\ba
T^{ab} &\equiv& \frac{1}{6\sqrt{2}}\,g \,\, 
\rme^{-\frac{2}{\sqrt{6}}\,\phi}\,(\G_{45})^{ab},\;\; 
h_{\m\n}^{ab} \equiv \rme^{-\frac{4}{\sqrt{6}}\phi}\,\Omega^{ab}\,
\cF_{\m\n},\;\;
H_{\mu\nu}^{ab} \equiv \rme^{\frac{2}{\sqrt{6}}\phi}\,
F_{\mu\nu}^I\,(\Gamma_I)^{ab}\,\,\,.
\label{hdef}
\ea
The gauge-covariant derivative ${\cal D}_\m$ acting on the Killing spinor is 
\beq
{\cal D}_\m\,\e_a = \nabla_\m\,\e_a +\fr12\,
g\,A^I_\m\,(\Gamma_{I\,45})_a^{\;\;b}\,\e_b \,\,\, ,
\eeq
with
\beq
\nabla_\mu \e_a \equiv \left( \partial_\mu+\frac{1}{4} \, 
\omega^{\, \, \, \, \alpha \beta}_\mu  \,
      \gamma_{\alpha \beta} \right) \, \e_a \,\,\, ,
\eeq
where $\omega^{\,\,\,\, \alpha \beta}_\mu$ is the
spin connection. Indices $\a, \b$ are tangent space (or flat) indices,
while $\m, \n$ are spacetime (or curved) indices. The 
$\gamma_{\alpha\beta\cdots}$ are the five-dimensional Dirac
matrices,
\[ \gamma_{\a_1 \cdots\a_n}=\frac{1}{n\,!} \,
\gamma_{[\a_1}\,\cdots\,\gamma_{\a_n]},\;\;\;\;\;\; n = 1, \cdots, 5 \,\,\, .\]
The equations of motion of the Lagrangian~(\ref{romans5})
are
\ba
R_{\m\n} &=& 2\,\pa_\m\f\,\pa_\n\f - 2\,{\rm e}^{-\frac{8}{\sqrt{6}}\f}\,
(\cF_\m^{\;\;\r}\,\cF_{\n\r} - \fr16\,g_{\m\n}\,\cF_{\r\s}\,\cF^{\r\s})\nn\\ 
& & - 2\,{\rm e}^{\frac{4}{\sqrt{6}}\f}\,(F_\m^{I\;\r}\,F^I_{\n\r} -
\fr16\,g_{\m\n}\,F_{\r\s}^I\,F^{I\;\r\s}) + 
\fr16\,g_{\m\n}\,g^2\,\rme^{-\frac{4}{\sqrt{6}}\f} \,\,\, ,
\label{einsteineq}\\
\Box\f &=& -\fr1{2\sqrt{6}}\,g^2\,{\rm e}^{-\frac{4}{\sqrt{6}}\f} 
+\fr2{\sqrt{6}}\,{\rm e}^{-\frac{8}{\sqrt{6}}\f}\,\cF^{\m\n}\,\cF_{\m\n} 
- \fr1{\sqrt{6}}\,{\rm e}^{\frac{4}{\sqrt{6}}\f}\,F^{I\;\m\n}\,F_{\m\n}^I \,\,\, ,
\label{scalareq}\\
{\cal D}_\n\,({\rm e}^{-\frac{8}{\sqrt{6}}\f}\,\cF^{\n\m}) &=&
\fr14\,e\,\varepsilon^{\m\n\r\s\t}\,F^I_{\n\r}\,F^I_{\s\t} \,\,\, ,
\label{abelianeq}\\
{\cal D}_\n\,({\rm e}^{\frac{4}{\sqrt{6}}\f}\,F^{I\;\n\m}) &=&
\fr12\,e\,\varepsilon^{\m\n\r\s\t}\,F^I_{\n\r}\,\cF_{\s\t} \,\,\, .
\label{nonabelian}
\ea
After some appropriate rescalings of the fields, the 
Lagrangian~(\ref{romans5}) can be written as
\ba
e^{-1}\,{\cal L} &=& R -\frac{1}{2} \, 
(\partial^\mu\phi) \, (\partial_\mu\phi) - \frac{1}{4} \, 
\rme^{-\frac{2}{\sqrt{6}}\f}\,F^{I \, \mu\nu} \, F^I_{\mu\nu}
- \frac{1}{4}\,\rme^{\frac{4}{\sqrt{6}}\f}\,\cF_{\m\n}\,\cF^{\m\n}\nn\\
& & + 4\,g^2\,\rme^{\frac{2}{\sqrt{6}}\f}
- \frac{1}{8} \,e^{-1}\,\varepsilon^{\m\n\r\s\t}\,F^I_{\r\s} \, F^I_{\t\k}\,
\cF_\t\,\,\, .
\label{canod5}
\ea
The Eq.~(\ref{canod5}) is the Lagrangian presented in reference
\cite{Cowdall:1999rs} with $G_2^{(1)} = G_2^{(2)} = 0$.

~


\section{Obtaining five-dimensional Romans' theory}


It was shown in~\cite{Lu:2000bw} that the five-dimensional 
$SU(2)\times U(1)$ gauged ${\cal N}=4$ supergravity can be obtained 
from reduction of type IIB supergravity on $S^5$. For our purpose, we
are interested in getting the five-dimensional gauged supergravity 
without the $U(1)$ gauge coupling. Turning off the $U(1)$ gauge
coupling can be thought of as taking a singular limit of $S^5$. In
this limit $S^5$ is deformed to $S^3\times T^2$. Following
this observation, instead of taking the singular limit of the metric and
field strength presented in~\cite{Lu:2000bw}, we will show that the
$SU(2)$ gauged supergravity in five dimensions can be derived from type
IIB on $S^3\times T^2$.
   
Let us begin with a subset of the bosonic sector of the ten-dimensional 
type IIB supergravity
\be
{\cal L}_{10} = \hat{R}\hat{*}\oneone - 
\fr12\hat{*}d\hat{\phi}\wedge d\hat{\phi} - 
\fr12\,\rme^{-\hat{\phi}}\,\hat{*}\hat{F}_3\wedge\hat{F}_3.
\label{10d}
\ee
The equations of motion of the ten-dimensional theory are
\ba
d(\rme^{-\hat{\phi}}\hat{*}\hat{F}_3) &=& 0\,,\nn\\
d(\hat{*}d\hat{\phi}) &=&
-\frac{1}{2}\,\rme^{-\hat{\phi}}\,\hat{*}\hat{F}_3\wedge
\hat{F}_3\,,\nn\\
\hat{R}_{\m\n}&=&\frac{1}{2}\pa_\m\hat{\phi}\pa_\n\hat{\phi} + 
\frac{1}{2}\rme^{-\hat{\phi}}\left[\hat{F}_{\m\r\s}\hat{F}_{\n}^{\;\;\r\s}
- \frac{1}{12}\hat{g}_{\m\n}\,\hat{F}_{\r\s\t}\hat{F}^{\r\s\t}\right]\,.
\label{10deq}
\ea
Following the procedure in~\cite{Cvetic:2000dm}, we reduce the 
ten-dimensional theory on $S^3$ and retain only $SU(2)$ subgroup of a 
full $SO(4)$ isometry group of $S^3$. The ans\"atze for the metric, the
scalar and the three form field are
\ba
d\hat{s}_{10}^2 &=&
\rme^{\frac{3}{\sqrt{10}}\,\tilde{\phi}}\,d\tilde{s}_7^2 + 
\frac{1}{4g^2}\,\rme^{-\frac{5}{\sqrt{10}}\,\tilde{\phi}}\,
\sum_{i=1}^3\,(\sigma^i - g\,A^i_1)^2,\nn\\
\hat{F}_3 &=& \tilde{F}_3 - \frac{1}{24g^2}\,\epsilon_{ijk}\tilde{h}^i\wedge
\tilde{h}^j\wedge\tilde{h}^k + 
\frac{1}{4g}\,\tilde{F}^i_2\wedge\tilde{h}^i,\nn\\
\hat{\phi} &=& \sqrt{10}\,\tilde{\phi},\nn\\
\tilde{h}^i &=& \sigma^i - g\,\tilde{A}^i\,,
\label{ansatz1}
\ea
where $\tilde{F}_3 = d\tilde{A}_2 + \frac{1}{4}\,\tilde{F}^i_2\wedge 
\tilde{A}^i_1 -
\frac{1}{24}g\,\epsilon_{ijk}\tilde{A}^i_1\wedge\tilde{A}^j_1\wedge
\tilde{A}^k_1$. Substituting the ans\"atze~(\ref{ansatz1}) into 
Eq.~(\ref{10deq}),
we obtain
\ba
d\,\tilde{F}_3 &=&
\frac{1}{4}\,\tilde{F}_2^i\wedge\tilde{F}_2^i\,,\;\;\;
d(\rme^{-\frac{4}{\sqrt{10}}\tilde{\phi}}\,\tilde{*}\tilde{F}_3) =
0\,,\nn\\
{\cal
D}(\rme^{-\frac{2}{\sqrt{10}}\tilde{\phi}}\,\tilde{*}\tilde{F}^i_2)
&=& \frac{1}{2}\,\rme^{-\frac{4}{\sqrt{10}}\tilde{\phi}}\,
\tilde{*}\tilde{F}_3\wedge\tilde{F}_2^i\,,\nn\\
d(\tilde{*}d\tilde{\phi}) &=& -\frac{2}{\sqrt{10}}\,
\rme^{-\frac{4}{\sqrt{10}}\tilde{\phi}}\,\tilde{*}\tilde{F}_3\wedge
\tilde{F}_3 - \frac{1}{\sqrt{10}}\,
\rme^{-\frac{2}{\sqrt{10}}\tilde{\phi}}\,\tilde{*}\tilde{F}^i_2\wedge
\tilde{F}^i_2\nn\\
& & - \frac{8}{\sqrt{10}}g^2\,\rme^{\frac{2}{\sqrt{10}}\tilde{\phi}}
\tilde{*}\oneone\,,\nn\\
\tilde{R}_{\m\n} &=& \frac{1}{2}\pa_\m\tilde{\phi}\pa_\n\tilde{\phi} + 
\frac{1}{4}\,\rme^{-\frac{4}{\sqrt{10}}\tilde{\phi}}\,\left[
\tilde{F}_{\m\r\s}\tilde{F}_{\n}^{\;\;\r\s} - 
\frac{1}{9}\tilde{g}_{\m\n}\,\tilde{F}_{\r\s\t}\tilde{F}^{\r\s\t}\right]\nn\\
& & \frac{1}{2}\,\rme^{-\frac{2}{\sqrt{10}}\tilde{\phi}}\,\left[
\tilde{F}_{\m\r}\tilde{F}_{\n}^{\;\;\r} - 
\frac{1}{9}\tilde{g}_{\m\n}\,\tilde{F}_{\r\s}\tilde{F}^{\r\s}\right] -
\frac{2}{3}g^2\,\rme^{\frac{2}{\sqrt{10}}\tilde{\phi}}\,\tilde{g}_{\m\n}\,.
\label{7deq1}
\ea  
Using odd-dimensional dualization~\cite{Townsend:1984xs} 
we change 3-form to 4-form
\be
\tilde{F}_4 =
\rme^{-\frac{4}{\sqrt{10}}\tilde{\phi}}\,\tilde{*}\,\tilde{F}_3\,,
\;{\rm or}\;\;\tilde{F}_3 =
- \rme^{\frac{4}{\sqrt{10}}\tilde{\phi}}\,\tilde{*}\,\tilde{F}_4 \, .
\label{dualization}
\ee
In terms of the 4-form field strength, the Eqs.~(\ref{7deq1}) become
\ba
d(\rme^{\frac{4}{\sqrt{10}}\tilde{\phi}}\,\tilde{*}\tilde{F}_4) &=&
\frac{1}{4}\,\tilde{F}_2^i\wedge\tilde{F}_2^i\,,\nn\\
{\cal
D}(\rme^{-\frac{2}{\sqrt{10}}\tilde{\phi}}\,\tilde{*}\tilde{F}^i_2)
&=&
\frac{1}{2}\,\tilde{F}_4\wedge\tilde{F}_2^i\,,\nn\\
d(\tilde{*}d\tilde{\phi}) &=& \frac{2}{\sqrt{10}}\,
\rme^{\frac{4}{\sqrt{10}}\tilde{\phi}}\,\tilde{*}\tilde{F}_4\wedge
\tilde{F}_4 - \frac{1}{\sqrt{10}}\,
\rme^{-\frac{2}{\sqrt{10}}\tilde{\phi}}\,\tilde{*}\tilde{F}^i_2\wedge
\tilde{F}^i_2\nn\\
& & - \frac{8}{\sqrt{10}}g^2\,\rme^{\frac{2}{\sqrt{10}}\tilde{\phi}}
\tilde{*}\oneone\,.
\label{7deq2}
\ea  
Eqs.~(\ref{7deq2}) together with Einstein's equations (for
simplicity, we do not write them down in~(\ref{7deq2})) constitute the 
equations of motion derived from the Lagrangian of $SU(2)$
gauged ${\cal N}=2$ supergravity in seven dimensions without
topological mass term~\cite{Townsend:1983kk,Salam:1983fa}
\ba
{\cal L}_7 &=& \tilde{R}\tilde{*}\oneone -
\fr12\,\tilde{*}d\tilde{\phi}\wedge d\tilde{\phi} -
\fr12\,\rme^{\frac{4}{\sqrt{10}}\tilde{\phi}}\,\tilde{*}\tilde{F}_4\wedge
\tilde{F}_4 - \fr12\,\rme^{-\frac{2}{\sqrt{10}}\tilde{\phi}}\,
\tilde{*}\tilde{F}^i_2\wedge \tilde{F}^i_2 \nn\\
& & + 4\,g^2\,\rme^{\frac{2}{\sqrt{10}}\tilde{\phi}}\,\tilde{*}\oneone +
\frac{1}{4}\,\tilde{F}_2^i\wedge\tilde{F}_2^i\wedge\tilde{A}_3\,, 
\label{7deq3}
\ea
where
\be
\tilde{F}_4 = d\tilde{A}_3\,,\;\;{\rm and}\;\tilde{F}_2^i =
d\tilde{A}_1^i +
\frac{1}{2}\,\epsilon_{ijk}\,\tilde{A}_1^j\wedge\tilde{A}_1^k.
\ee
The above seven-dimensional gauged supergravity whose Lagrangian is 
Eq.~(\ref{7deq3}) can also be obtained from an appropriate
truncation of a gauged supergravity derived from reducing type IIA
supergravity on $S^3$~\cite{Cvetic:2000ah}.
Having obtained the seven-dimensional gauged supergravity, we
reduce it on $T^2$ following~
\cite{Giani:1984dw,Cowdall:1998fn,Cowdall:1999rs}. The ansatz for
reduction of the seven-dimensional gauged supergravity on $T^2$ is 
\ba
ds_7^2 &=& \rme^{\frac{4}{5\sqrt{6}}\phi}\,ds_5^2
+\rme^{-\frac{6}{5\sqrt{6}}\phi}\,(dY^2 + dZ^2)\,,\nn\\
\tilde{F}_3 &=& \cF_2\wedge dZ\wedge dY\,,\nn\\
\tilde{F}_2^i &=& F^i_2\,.
\ea
The five-dimensional Lagrangian obtained from this process is
\begin{eqnarray}
{\cal L}_5 & = & R*\oneone - \fr12\,*d\f\wedge d\f - 
\fr12\,\rme^{-\frac{2}{\sqrt{6}}\f}\,*F^i_2\wedge F^i_2  
-\fr1{2}\,\rme^{\frac{4}{\sqrt{6}}\f}\,*\cF_2\wedge \cF_2\nn\\
& &  + \frac{1}{4}\,F^i_2\wedge F^i_2\wedge \cA_1 \, + 
4\,g^2\,\rme^{\frac{2}{\sqrt{6}}\f}*\oneone \,\,\, , 
\label{5dlag}
\end{eqnarray}
where
\be
\cF_2 = d\,\cA_1\;,\;\;\;{\rm and}\;\;\;F^i_2 = dA_1^i +
\frac{1}{2}\,\epsilon_{ijk}\, A^j_1\wedge A^k_1\,\,\,.
\ee
The Lagrangian~(\ref{5dlag}) is the $SU(2)$ gauged ${\cal N}=4$
supergravity with vanishing $U(1)$ coupling constant and without two 2-form
potentials. Eq.~(\ref{5dlag}) is written in terms of canonical
normalized scalar $\phi$ and the signature of the space time is mostly plus.

The complete reduction ansatz from 10 to 5 dimensions is
\ba
d\hat{s}_{10}^2 &=& \rme^{\frac{13}{5\sqrt{6}}\f}\,ds_5^2 + 
\rme^{\frac{3}{5\sqrt{6}}\f}\,(dY^2 + dZ^2) + 
\frac{1}{4g^2}\,\rme^{-\frac{3}{\sqrt{6}}\f}\,\sum_{i=1}^3(\s^i -
g\,A_1^i)^2\,,\nn\\
\hat{F}_3 &=&  \rme^{\frac{4}{\sqrt{6}}\f}*\cF_2 -
\frac{1}{24g^2}\,\epsilon_{ijk}\, h^i\wedge h^j\wedge h^k + 
\frac{1}{4}\,\sum_{i=1}^3\,F_2^i\wedge h^i\,,\nn\\
\hat{\phi} &=& \sqrt{6}\,\phi\;,\;\;\;{\rm and}\;\;\;h^i = \s^i - g\,A_1^i\,.
\label{typeIIBansatz}
\ea
The ansatz~(\ref{typeIIBansatz}) tells us that any solution of the
five-dimensional gauged supergravity can be up-lifted to ten
dimensions and, this is a solution corresponding to the NS-fivebrane.
It is not clear that a supergravity obtained from reducing type IIA
on $S^3$ is dual to a theory obtained from reducing type IIB on $S^3$
in the same sense as T-duality. However, for a particular subset of
type IIA and type IIB, we will show that type IIA on $S^1\times S^3$ is
equivalent to type IIB on $S^3\times S^1$. Therefore, any solution of
the six-dimensional and the five-dimensional gauged supergravities can be 
uplifted to either type IIA or type IIB supergravities.

The reduction of a subset of type IIB supergravity on 
$S^3\times S^1$ is presented above. It is not hard to see that
reducing the seven-dimensional theory whose Lagrangian is
Eq.~(\ref{7deq3}) on a circle produces a subset of Romans' theory in
six dimensions. On the other hand, the same subset of the
Romans' theory in six dimensions was obtained by reducing type IIA on
$S^1\times S^3$~\cite{Nunez:2001pt}, together with 
the dualization of the 3-form field. 
The Lagrangian and equations of motion
of the six-dimensional theory~\cite{Nunez:2001pt} after dualizing the
3-form field are
\begin{eqnarray}
{\cal L}_6 &=& \bar{R} - 
\frac{1}{2}(\partial\bar{\phi})^2 - 
\frac{1}{2}{\rm e}^{-\frac{1}{\sqrt{2}}\bar{\phi}}\,\bar{*}
\bar{F}^i_2\wedge\bar{F}^i_2 +
4\,g^2 {\rm
e}^{\frac{1}{\sqrt{2}}\bar{\phi}}\,\bar{*}\oneone\nn\\
& & +\frac{1}{4}\,\bar{F}^i_2\wedge \bar{F}_2^i\wedge \bar{A}_2
- \frac{1}{2}{\rm e}^{\sqrt{2}\bar{\phi}}\bar{F}_3\wedge \bar{F}_3\,\, ,\nn\\
d(\rme^{\sqrt{2}\bar{\phi}}\,\bar{*}\bar{F}_3) &=& 
\frac{1}{4}\,\sum_{i=1}^3\,\bar{F}^i_2\wedge
\bar{F}_2^i \,\, ,\label{6dimeq}\\
{\cal D}(\rme^{-\frac{1}{\sqrt{2}}\bar{\phi}}\,\bar{*}\bar{F}_2^i) &=& 
\frac{1}{2}\bar{F}_3\wedge \bar{F}_2^i \,\,\, ,\nonumber\\
d(\bar{\phi}\,\bar{*}d\bar{\phi}) &=& -\frac{1}{\sqrt{2}}
\rme^{\sqrt{2}\bar{\phi}}
\,\bar{*}\bar{F}_3\wedge \bar{F}_3 +
\frac{1}{2\sqrt{2}}\,\rme^{-\frac{1}{\sqrt{2}}\bar{\phi}}\,\sum_{i=1}^3\, 
\bar{*}\bar{F}^i_2\wedge \bar{F}^i_2 + \frac{4}{\sqrt{2}}\,g^2
\,\rme^{\frac{1}{\sqrt{2}}
\bar{\phi}}\,\bar{*}\oneone \,\, ,\nn
\end{eqnarray}
where $\bar{F}_3 = d \bar{A}_2$ and $\bar{F}^i_2 = d\bar{A}_1^i +
\frac{1}{2}\epsilon_{ijk}\,\bar{A}^j_1\wedge \bar{A}_1^k$. Reducing
the above six-dimensional theory on a circle gives the
five-dimensional theory without $U(1)$ gauged coupling. The ansatz of
reduction from type IIA supergravity on $S^1\times S^3\times S^1$ is
\begin{eqnarray}
d\hat{s}_{10}^2 &=& {\rm e}^{\frac{7}{8\sqrt{6}}\phi}\, ds_5^2 + 
\frac{1}{4g^2}\,{\rm e}^{-\frac{9}{8\sqrt{6}}\phi}\,
\sum_{i=1}^3\,\left(\sigma^i - g\,A^i_1\right)^2 + 
{\rm e}^{\frac{15}{8\sqrt{6}}\phi}\,dY^2 +
\rme^{-\frac{9}{8\sqrt{6}}\phi}\,dZ^2 \,\,\, , \nonumber\\
\hat{F}_4 &=& \left(\,\rme^{\frac{4}{\sqrt{6}}\f}\,*\cF_2 -
\frac{1}{24g^2}\epsilon_{ijk}\,h^i\wedge h^j\wedge h^k + 
\frac{1}{4g}\,F^{i}_2\wedge h^i\,\right)\wedge dY 
\,\,\, ,\nonumber\\
\hat{\phi\baselineskip=20pt plus 1pt minus 1pt
} &=& \frac{3}{4\sqrt{6}}\,\phi \,\,\, .
\label{metricansatz}
\ea


\section{Duals of 3-dimensional ${\cal{N}}=1$ SYM 
theory on a torus}


In this section we study the supergravity dual of 
a 3-dimensional ${\cal{N}}=1$ SYM theory on a torus. 
The gravitational system we are dealing with can be understood as follows. 
Let us consider $N$ type IIB NS-fivebranes. If the fivebranes were flat,
the isometries of this system would be $SO(1,5) \times SO(4)$. 
The first corresponds to the Lorentz group on the flat fivebrane worldvolumes, 
while the second one is the corresponding rotation group of the $S^3$ 
tranverse
to the fivebrane directions. Since the NS-fivebranes are not flat but wrapped  
on a second $S^3$ (in the five-dimensional Romans' theory), 
we have the following chain of breaking of the isometries
$SO(1,9) \rightarrow SO(1,5) \times SO(4) \rightarrow SO(4) \times SO(4)$. There
is also an additional isometry group corresponding to the torus, where the two
additional spatial directions of the fivebrane are wrapped. 
On the other hand, 
the supergravity solution that we consider here has a global $SO(4)$ symmetry, and
its corresponding ansatz for the five-dimensional metric has the 
$R^1 \times S^3 \times R^1$ geometry.
The $R^1$'s correspond to the time and the radial coordinate, respectively.
In ten dimensions, the solution has the geometry of the form 
$(R^1_0 \times S^3_{1,2,3} \times R^1_4) \times T^2_{5,6} \times S^3_{7,8,9}$, 
where the lower indices label the coordinates.
Recall from the previous section that the seven-dimensional
supergravity is related to the five-dimensional one through a
$T^2$ reduction, whereas the up-lifting to 10-dimensional theory is 
obtained through an $S^3$.
In the table below, we schematically show the global
structure of the ten-dimensional metric. The first five coordinates
are arbitrarily chosen to represent the five-dimensional metric for 
the Romans' theory. 
\begin{center}
\begin{tabular}{|c|c|c|c|c|c|}
\hline 

 0 & 1 \, 2 \, 3 & 4 & 5 \, 6 & 7 \, 8 \, 9   \\ 
\hline 


 $R^1_0$ & $S^3_{1,2,3}$ & $R^1_4$ & $T^2_{5,6}$ & $S^3_{7,8,9}$  \\
\hline
\end{tabular}
\end{center}
From the above table, one can see that the NS-fivebrane is wrapped on 
the $S^3$ (which belongs to the five-dimensional
$SU(2)$ gauged supergravity metric ansatz), while 
its other two spatial directions are wrapped on $T^2_{5,6}$,
{\it i.e.}, the directions 5 and 6. 

Now, we focus on the twisting preserving two supercharges. 
As already mentioned above, there are three spatial directions of the 
NS-fivebranes wrapped on $S^3$.
Therefore, the supersymmetry will be realized through a twisting.
Also notice that the NS-fivebranes have two directions on a torus, so that
these are not involved in a twisting. The brane worldvolume is on $R^1 \times S^3 \times T^2$.
The non-trivial part of the spin connection on this worlvolume is the $SU(2)$ connection on the
spin bundle of $S^3$. On the other hand, the normal bundle to the NS-fivebrane in the $G_2$
manifold is given by $SU(2) \times SU(2)$, one of them being the 
spin bundle of $S^3$.
In this case, the twisting consists in the indentification of the $SU(2)$ group of the spin bundle
with one of the factors in the R-symmetry group of the fivebrane, {\it i.e.} 
$SO(4)_R \rightarrow SU(2)_L \times SU(2)_R$.
It leads to a diagonal group $SU(2)_D$, so that it gives a twisted
gauge theory.
The resulting symmetry group is $SO(1,2) \times SU(2)_D \times SU(2)_R$.
In the UV limit the global symmetry is $SO(1,5) \times SU(2)_L \times SU(2)_R$,
so that the four scalars transform as the representation $({\bf 1}, {\bf 2}, {\bf 2})$
and there are also 16 supercharges. After the twisting 
we get 2 fermions (which are the
two supercharges of the remaining unbroken supersymmetry) 
transforming in the $({\bf 2}, {\bf 1}, {\bf 1})$ 
representation of $SO(1,2) \times SU(2)_D \times SU(2)_R$. There are
no scalars after twisting,
while we get one vector field as it is before the twisting.
Therefore, $1/8$ of the supersymmetries are preserved, which is
related to the fact that the two $S^3$'s, together with the radial 
coordinate are embedded in a $G_2$ manifold.
In this way, since our IR limit corresponds to set the radial coordinate 
to be zero, it implies (as we will see) that the  $S^3$ part of the
fivebrane will reduce to a point. This is in contrast with the fact that
the transverse $S^3$ and the torus get fixed radii.
It shows that when one moves to the IR 
of the gauge theory (flowing in the radial coordinate on the gravity dual) three of the dimensions
become very small  and no low energy massless modes are excited on this two-space.
Therefore, effectively far in the IR the gauge theory is three-dimensional.

In order to show explicitly how the theory flows to a 3-dimensional SYM theory on a torus,
we briefly describe the $SO(4)$-symmetric solution of the five-dimensional Romans'
supergravity presented in \cite{Chamseddine:2001hk}. Following that reference,
let us consider a static field configuration, invariant under the $SO(4)$ global symmetry
group of spatial rotations. As we already mentioned, the metric ansatz has the structure
$R \times S^3 \times R$ and it can be written as 
\be
d s^2_5 = e^{2 \nu(r)} \, dt^2 - \frac{1}{M(r)} \, dr^2 - r^2 \, d \Omega^2_3 \,\,\, , 
\label{metricfive}
\ee
where $d \Omega^2_3$ is the metric on $S^3$. 
Notice that here we have adopted the mostly minus signature. 
In order to define the field configurations on this geometry it is useful to introduce
the left-invariant forms $\s^i$ on $S^3$, such that they satisfy the Maurer-Cartan equation
\be
d \s^i + \epsilon_{i j k} \,\, \s^j \, \wedge \, \s^k = 0 \,\,\, ,
\ee
and consequently $d \Omega^2_3 = \s^i \,\, \s^i$. We consider the non-Abelian gauge potential
components written in terms of the left-invariant forms
\be
A^i = A^i _\m \, d x^\m = [w(r)+1] \, \s^i \,\,\, ,
\ee
so that they are invariant under the combined action of the $SO(4)$ rotations  and the $SU(2)$
gauge transformations. The corresponding field strength is purely
magnetic and is given by
\be
F^i = dw \, \wedge \, \s^i + \frac{1}{2} \,  [w(r)^2-1] \, \epsilon_{i j k} \,\, 
\s^j \, \wedge \, \s^k \,\,\, ,
\ee
while for the Abelian gauge potential we consider a purely electric ansatz
\be
f(r) = Q(r) \, dt \, \wedge \, dr \,\,\, .
\ee
All the rest of functions, {\it i.e.} $\n$, $M$, $w$, $Q$ and the dilaton $\phi$ 
are only dependent of the radial coordinate $r$. From the equation of motion of the
dilaton the relation
\be
\n(r) = \sqrt{\frac{2}{3}} \, (\phi(r) - \phi_0) 
\ee
can be obtained,
where $\phi_0$ is an integration constant. On the other hand, from the equation of motion
of the Abelian field, the following result
\be
Q(r) = \frac{e^{5 \n(r)}}{\sqrt{M(r)}\, r^3} \, [2 \, w(r)^3 - 6 \, w(r) + H] 
\ee
is derived,
where $H$ is an integration constant. The rest of equations of motion with the field
configuration and the metric ansatz described above, can be found in reference
\cite{Chamseddine:2001hk}. The solution must satisfy the equations obtained by setting to
zero the supersymmetry transformations for gauginos and gravitinos Eqs.(\ref{gautransf}) and
(\ref{gratransf}), such 
that the following first order differential equations are obtained
\ba
M(r) & = & \left( \frac{1}{3} \, \zeta^2 \, V - w  \right)^2 +
2 \, \zeta^2\, (w^2-1)^2 - \frac{2}{3} \, (w^2 -1)+ \frac{1}{18 \zeta^2} \,\,\, , \nn \\
\frac{d w(r)}{d \log{r}} & = & \frac{1}{6 \, \zeta^2 \, M} \left\{ 
- 2 \, V \,  (w^2 -1) \, \zeta^4 + (H - 4 \, w^3) \,  \zeta^2 - w
\right\} \,\,\, , \nn \\
\frac{d \zeta(r)}{d \log{r}} & = & -\frac{\zeta}{3 \, M} \times \nn \\
& &  \left\{ 
V^2 \, \zeta^4 + 12 \, \zeta^2 \, (w^2-1)^2 - 4 \, V \, w \, \zeta^2 +
w^2 +2
\right\} \,\,\, , \label{BOGO}
\ea
where we have defined $\zeta(r) = \exp[\n]/r$ and $V(r) = 2 \, w(r)^3 - 6 \, w(r) +H$. These
equations are compatible with the equations of motion derived from the Romans' five dimensional
Lagrangian given in section 2, and any solution of these first order
differential equations preserves two supersymmetries.  

Since we are interested in the IR limit, {\it i.e.} when $r \rightarrow 0$, we obtain the expansions
of the functions defining the metric, the magnetic non-Abelian and the electric Abelian
fields for the five-dimensional ansatz. They are
\ba
w(r) & = & 1 - \frac{1}{24} \, r^2 + \cdot \, \cdot \,  \cdot  \,\,\, , \nn \\
\zeta(r) & = & \frac{1}{r} + \frac{7}{288} \, r + \cdot \, \cdot \,  \cdot   , \nn \\
M(r) & = & 1 + \frac{5}{144} \, r^2 + \cdot \, \cdot \,  \cdot
\,\,\, , 
\ea
and straightforwardly
\be
\n(r)  =  \frac{7}{288} \, r^2 + \cdot \, \cdot \,  \cdot \,\,\, ,
\ee
while for the dilaton we obtain
\be
\phi(r)  =  \phi_0 +\frac{7}{288} \, \sqrt{\frac{3}{2}} \, \, r^2 + \cdot \, \cdot \,  \cdot    \,\,\, . 
\ee
In this case we have taken $H$ to be 4. Also, we get
\be
Q(r)=\frac{1}{96} \, r+ \frac{13}{13824} \, r^3+ \cdot \cdot \cdot \, . 
\ee
In this way, one can see that in the
IR limit the non-Abelian gauge potential has a core, while both field strengths, {\it i.e.}, the
Abelian and the non-Abelian one are of the order $r$ around $r=0$. 
Futhermore, the above solution can be up-lifted, following the ans\"atze 
presented in the section 3, to either type IIA or type IIB theories.  
In these two cases, the IR limit turns out to be the same. \\

~
~

{\it Up-lifting to type IIB theory}

~

Firstly, we consider the case when the solution
is up-lifted to type IIB supergravity. 
From Eq.(\ref{typeIIBansatz}) the 10-dimensional metric is
\ba
d\hat{s}_{10}^2 &=& - \rme^{\frac{13}{5\sqrt{6}}\f}\,
\left( \rme^{2 \nu(r)} \, dt^2 - \frac{1}{M(r)} \, dr^2 - r^2 
\, d \Omega^2_3 \right) \nn \\
& & +  
\rme^{\frac{3}{5\sqrt{6}}\f}\,(dY^2 + dZ^2) + 
\frac{1}{4g^2}\,\rme^{-\frac{3}{\sqrt{6}}\f}\,\sum_{i=1}^3(\s^i -
g\,A_1^i)^2 \,\,\, , \nn\\
\hat{\phi} &=& \sqrt{6}\,\phi\, .
\label{metricfinalIIB}
\ea
Therefore, using the previously calculated IR expansion  
we can obtain the radii of the different manifolds. Thus,
for the $S^3$ involving the coordinates 1, 2, and 3,
the radius is given by 
\be
R^2_{1,2,3} = \rme^{13 \phi_0/5 \sqrt{6}} \, r^2 + {\cal {O}}(r^4) \,\,\, ,
\label{r123iib}
\ee
so that we can see how the radius of the $S^3$ in the five-dimensional
metric ansatz shrinks to zero in the IR. On the other hand, the radii of 
$T^2_{5,6}$, $S^3_{7,8,9}$ remain finite as we can see as follows
\ba
R^2_{T} &=&\rme^{\frac{\sqrt{3}}{5\sqrt{2}}\f_0}\left(
1 - \frac{7}{960}\, r^2 + {\cal {O}}(r^3)\right) \,\,\, , \nn \\
R^2_{7,8,9} &=&
\frac{1}{4g^2}\rme^{-\frac{3}{\sqrt{6}}\phi_0}\,\left(
1 - \frac{21}{576}\, r^2 + {\cal {O}}(r^3)\right) \,\,\, .
\label{rtiib}
\ea
Without loss of generality, we can set $\phi_0$ to zero.
It is obvious from Eqs.~(\ref{r123iib}) and (\ref{rtiib}) that  
in the limit $r \rightarrow 0$, $R_T$ and $R_{7,8,9}$ remain finite, 
while $R_{1,2,3} \rightarrow 0$. Since the type IIB NS-fivebrane is 
wrapped on $S^3_{1,2,3}$, $T^2$,
and in the IR limit $S^3_{1,2,3}$ effectively reduces to a point, 
in this limit we obtain a twisted gauge field theory defined on the torus. \\

~
~

{\it Up-lifting to type IIA theory}

~

Now, we consider the metric given in Eq.(\ref{metricansatz}) 
\begin{eqnarray}
d\hat{s}_{10}^2 &=& - {\rm e}^{\frac{7}{8\sqrt{6}}\phi}\, 
\left(\rme^{2 \nu(r)} \, dt^2 - \frac{1}{M(r)} \, dr^2 - r^2 
\, d \Omega^2_3 \right)  \nn \\
& & +
\frac{1}{4g^2}\,{\rm e}^{-\frac{9}{8\sqrt{6}}\phi}\,
\sum_{i=1}^3\,\left(\sigma^i - g\,A^i_1\right)^2 + 
{\rm e}^{-\frac{9}{8\sqrt{6}}\phi}\,dZ^2 +
\rme^{\frac{15}{8\sqrt{6}}\phi}\,dY^2 \,\,\, , \nonumber\\
\hat{\phi\baselineskip=20pt plus 1pt minus 1pt
} &=& \frac{3}{4\sqrt{6}}\,\phi \,\,\, .
\label{metricfinal}
\ea
Therefore, as we did for the type IIB case,
we can obtain the radius
\be
R^2_{1,2,3} = {\rm e}^{\frac{7}{8\sqrt{6}}\f_0} \, r^2 + 
{\cal {O}}(r^4) \,\,\, ,
\ee
which shrinks to zero in the IR limit. In addition, the radii of 
$S^1_{5}$, $S^3_{6,7,8}$ and $S^1_9$ are finite 
\ba
R^2_{5} &=&{\rm e}^{\frac{15}{8\sqrt{6}}\f_0}\,\left(
1 + \frac{105}{4608}\, r^2 + {\cal {O}}(r^3)\right) \,\,\, , \nn \\
R^2_{6,7,8} &=& \frac{1}{4g^2}\,{\rm e}^{-\frac{9}{8\sqrt{6}}\phi_0}
\left( 1 - \frac{27}{4608}\, r^2 + {\cal {O}}(r^3)\right) \,\,\, , \nn \\
R^2_{9} &=& {\rm e}^{-\frac{9}{8\sqrt{6}}\f_0}\,\left(1 -
 \frac{27}{4608}\, r^2 + {\cal {O}}(r^3)\right) \,\,\, .
\ea
Again, by considering $\phi_0=0$, in the IR, the radii 
$R_{5}=R_{9}$ and $R_{6,7,8}=1/(2 g)$, while $R_{1,2,3} \rightarrow 0$.
Since the type IIA NS-fivebrane is wrapped on $S^3_{1,2,3}$, $S^1_5$ and $S^1_9$,
and in the IR limit $S^3_{1,2,3}$ effectively shrinks to a point as in the
type IIB case, we get the same geometric reduction as in the previous case.
Note that this can be obtained when $\phi_0=0$, so that the radii of the
torus (in type IIB case) and the two $S^1$'s (in type IIA case) are exactly the same.

In addition, in both cases one can use the criterion for confinement given in references
\cite{Kinar:2000vq,Sonnenschein:1999if}, in order to show that the corresponding
static potential is confining.

~
~

{\it The singular $SO(4)$-symmetric solution}

~

A solution with no electric Abelian fields can
be obtained by setting $H$ to zero. It implies that
$w$, $V$ and also $Q$ are zeros, as we expected since
no electric field are excited. In this way, the
first order differential equations (\ref{BOGO})
can be easily integrated, yielding the relation
\be
r=r_0 \, \frac{{\rm e}^{1/24 \zeta^2}}{\sqrt{\zeta}} \,\,\, ,
\ee
where $r_0$ is an integration constant. The metric is given by 
\be
d s^2_5 = r_0^2 \,\, {\rm e}^{1/(12 \zeta^2)} \, \left(
\zeta \, dt^2 - \frac{1}{8 \, \zeta^5} \, d \zeta^2 -
\frac{1}{\zeta} \, d\Omega^2_3 \right) \,\,\, .
\ee
In adition, for the dilaton we have the following relation
\be
{\rm e}^{\sqrt{2} \phi/\sqrt{3}} = r_0 \,\, {\rm e}^{1/24 \zeta^2} \,\, \sqrt{\zeta}
\,\,\, .
\ee
Using the criterion of reference \cite{Maldacena:2001mw} it is straightforward to see
that the IR singularity is not acceptable, both in type IIA and type IIB theories.

~


\section{Discussion}


From the point of view of the supergravity theories, the $SO(4)$-symmetric solution
of the Romans' five-dimensional theory can be up-lifted to either type IIB 
or type IIA supergravities. These are obtained through the up-lifting to seven
and six-dimensional supergravities, respectively. It means that the 10-dimensional
system consists of NS fivebranes, either type IIB or IIA. On the other hand, in our previous paper
\cite{Nunez:2001pt}, we constructed a non-Abelian solution that is 
identical to the solution
obtained in references \cite{Chamseddine:1997nm,Chamseddine:1998mc}. It 
has been interpreted in \cite{Maldacena:2001yy} as a wrapped NS-fivebrane.
In this case, it was a gravity dual of a theory very similar to ${\cal {N}}=1$ super Yang-Mills.
In addition, in \cite{Nunez:2001pt} we interpreted the massless solution of the Romans'
six-dimensional theory as the same NS-fivebrane with a compactified direction.
In such a situation, the IR theory was ${\cal {N}}=2$ SYM in three dimensions. In these
cases, the fact that their actions in the string frame are similar, for the massless six-dimensional
supergravity and the corresponding seven-dimensional one with vanishing topological mass, indicates
that those are the same system. We can see a similar issue in the five-dimensional 
supergravity studied here, since again the 10-dimensional system involves NS-fivebranes. Thus, 
from the analysis in the present paper, we conclude that this is the natural extension of the
six and seven-dimensional results to five dimensions. In fact, in the IR limit this case corresponds
to ${\cal {N}}=1$ super Yang-Mills theory on a torus, which is confining.
This IR theory is interesting on its own. 
Although many aspects of three-dimensional super Yang-Mills 
theories have been considered\cite{Bergman:1999na,Dorey:2000rb,Aharony:1997},
some aspects of three-dimensional ${\cal {N}}=1$ super 
Yang-Mills theory on a torus are still poorly understood. Therefore, 
the results obtained here can be an interesting  motivation for
further studies since we have presented a gravity dual of 
${\cal {N}}=1$ super Yang-Mills theory on a torus.

We remark that
the singular solution obtained in \cite{Chamseddine:2001hk} is produced when the electric Abelian fields 
are turn off. This solution has a singular $g_{00}$ even when it is considered in the 
ten-dimensional theory. This means that this solution does not represent any gauge field
theory in the IR. We think that it would be interesting to know if the presence of electric
Abelian fields is related to a rotation of the fivebranes, leading to a desingularization
of the solution. Although, we think that this point deserves further investigation, we can discuss
here a little about this mechanism. The issue of the resolution of the singularity can be understood
as follows. We recall one of the cases studied in \cite{Nunez:2001pt}, which has been interpreted as 
the gravity dual of the three-dimensional ${\cal {N}}=2$ super Yang-Mills theory. Actually,
this solution is related to the one given in \cite{Maldacena:2001yy}, and it represents
a smeared NS-fivebrane on $S^2$ after T-duality. What is worth stressing is that the resolution
of the singularity in this case was produced by the excitation of non-Abelian fields. In the case
of the metric of Eq.(\ref{metricfive}), as we have seen, in order to obtain a non-singular metric
it is necessary to turn on the electric Abelian fields. We also have to recall that for that particular
case one forced the metric on the $S^3$ to be $r^2$. We think that this fact induces the singularity,
so that the non-Abelian fields are not enough in order to prevent it, as in the cases of 
\cite{Maldacena:2001yy,Nunez:2001pt}. It would be interesting to see what happens if instead of $r^2$ we
write a more general function of $r$. It would probably render a similar situation as in 
\cite{Maldacena:2001yy,Nunez:2001pt}, {\it i.e.} the resolution of the singularity with only
non-Abelian fields.


\newpage

\centerline{\bf Acknowledgements}

~

We are grateful to Carlos N\'u\~nez for collaborating in the early
stages of this project and for illuminating discussions.
We thank Amihay Hanany, Inyong Park, Christopher Pope and Matthew
Strassler for helpful discussions. M.S. would like to thank 
Ian Kogan and Fidel Schaposnik for correspondence. 
The work of M.S. is supported in part by funds provided by the 
U.S. Department of Energy under cooperative research 
agreement $\#$DF-FC02-94ER40818, CONICET of Argentina, 
Fundaci\'on Antorchas of Argentina and The British Council.

~

~


\begin{thebibliography}{99}

\bibitem{Maldacena:1998re}
J.~Maldacena,
``The large $N$ limit of superconformal field theories and supergravity,''
Adv.\ Theor.\ Math.\ Phys.\ {\bf 2} (1998) 231
[hep-th/9711200].

\bibitem{Gubser:1998bc}
S.~S.~Gubser, I.~R.~Klebanov and A.~M.~Polyakov,
``Gauge theory correlators from non-critical string theory,''
Phys.\ Lett.\ B {\bf 428} (1998) 105
[hep-th/9802109].

\bibitem{Witten:1998qj}
E.~Witten,
``Anti-de Sitter space and holography,''
Adv.\ Theor.\ Math.\ Phys.\ {\bf 2} (1998) 253
[hep-th/9802150].

\bibitem{Maldacena:2001mw}
J.~Maldacena and C.~N\'u\~nez,
``Supergravity description of field theories on curved manifolds and a no  go theorem,''
Int.\ J.\ Mod.\ Phys.\ A {\bf 16}, 822 (2001)
[hep-th/0007018].

\bibitem{Maldacena:2001yy}
J.~M.~Maldacena and C.~N\'u\~nez,
``Towards the large $N$ limit of pure ${\cal N} = 1$ super Yang-Mills,''
Phys.\ Rev.\ Lett.\ {\bf 86} (2001) 588
[hep-th/0008001].

\bibitem{Acharya:2000mu}
B.~S.~Acharya, J.~P.~Gauntlett and N.~Kim,
``Fivebranes wrapped on associative three-cycles,''
hep-th/0011190.

\bibitem{Nieder:2000kc}
H.~Nieder and Y.~Oz,
``Supergravity and D-branes wrapping special Lagrangian cycles,''
hep-th/0011288.

\bibitem{Gauntlett:2000ng}
J.~P.~Gauntlett, N.~Kim and D.~Waldram,
``M-fivebranes wrapped on supersymmetric cycles,''
hep-th/0012195.

\bibitem{Nunez:2001pt}
C.~N\'u\~nez, I.~Y.~Park, M.~Schvellinger and T.~A.~Tran,
``Supergravity duals of gauge theories from $F(4)$ gauged supergravity in  six dimensions,''
JHEP {\bf 0104}, 025 (2001)
[hep-th/0103080].

\bibitem{Edelstein:2001pu}
J.~D.~Edelstein and C.~Nunez,
``D6 branes and M-theory geometrical transitions from gauged  supergravity,''
JHEP {\bf 0104}, 028 (2001)
[hep-th/0103167].

\bibitem{Bershadsky:1996qy}
M.~Bershadsky, C.~Vafa and V.~Sadov,
``D-Branes and Topological Field Theories,''
Nucl.\ Phys.\ B {\bf 463} (1996) 420
[hep-th/9511222].


\bibitem{Itzhaki:1998dd}
N.~Itzhaki, J.~M.~Maldacena, J.~Sonnenschein and S.~Yankielowicz,
``Supergravity and the large $N$ limit of theories with sixteen 
supercharges,''
Phys.\ Rev.\ D {\bf 58}, 046004 (1998)
[hep-th/9802042].

\bibitem{Chamseddine:2001hk}
A.~H.~Chamseddine and M.~S.~Volkov,
``Non-Abelian vacua in $D = 5$, ${\cal N} = 4$ gauged supergravity,''
JHEP {\bf 0104}, 023 (2001)
[hep-th/0101202].

\bibitem{Romans:1986ps}
L.~J.~Romans,
``Gauged ${\cal N}=4$ Supergravities In Five-Dimensions And Their 
Magnetovac Backgrounds,''
Nucl.\ Phys.\ B {\bf 267}, 433 (1986).

\bibitem{Romans:1986tw}
L.~J.~Romans,
``The $F(4)$ Gauged Supergravity In Six-Dimensions,''
Nucl.\ Phys.\ B {\bf 269}, 691 (1986).
For more recent developments of $F(4)$ gauged
supergravity coupled to matter also see 
R.~D'Auria, S.~Ferrara and S.~Vaula,
``Matter coupled F(4) supergravity and the AdS(6)/CFT(5)  correspondence,''
JHEP {\bf 0010}, 013 (2000)
[hep-th/0006107],
L.~Andrianopoli, R.~D'Auria and S.~Vaula,
``Matter coupled F(4) gauged supergravity Lagrangian,''
JHEP {\bf 0105}, 065 (2001)
[hep-th/0104155].


\bibitem{Maldacena:2000dr}
J.~Maldacena and L.~Maoz,
``De-singularization by rotation,''
hep-th/0012025.

\bibitem{Cowdall:1999rs}
P.~M.~Cowdall,
``On gauged maximal supergravity in six dimensions,''
JHEP{\bf 9906}, 018 (1999)
[hep-th/9810041].

\bibitem{Lu:2000bw}
H.~Lu, C.~N.~Pope and T.~A.~Tran,
``Five-dimensional ${\cal N} = 4$, $SU(2)\times U(1)$ gauged 
supergravity from type IIB,''
Phys.\ Lett.\ B {\bf 475}, 261 (2000)
[hep-th/9909203].

\bibitem{Cvetic:2000dm}
M.~Cvetic, H.~Lu and C.~N.~Pope,
``Consistent Kaluza-Klein sphere reductions,''
Phys.\ Rev.\ D {\bf 62} (2000) 064028
[hep-th/0003286].

\bibitem{Townsend:1984xs}
P.~K.~Townsend, K.~Pilch and P.~van Nieuwenhuizen,
``Selfduality In Odd Dimensions,''
Phys.\ Lett.\  {\bf 136B}, 38 (1984)
[ {\bf 137B}, 443 (1984)].

\bibitem{Townsend:1983kk}
P.~K.~Townsend and P.~van Nieuwenhuizen,
``Gauged Seven-Dimensional Supergravity,''
Phys.\ Lett.\ B {\bf 125}, 41 (1983).

\bibitem{Salam:1983fa}
A.~Salam and E.~Sezgin,
``$SO(4)$ Gauging of ${\cal N}=2$ Supergravity In Seven-Dimensions,''
Phys.\ Lett.\ B {\bf 126}, 295 (1983).

\bibitem{Cvetic:2000ah}
M.~Cvetic, H.~Lu, C.~N.~Pope, A.~Sadrzadeh and T.~A.~Tran,
``$S^3$ and $S^4$ reductions of type IIA supergravity,''
Nucl.\ Phys.\ B {\bf 590}, 233 (2000)
[hep-th/0005137].

\bibitem{Giani:1984dw}
F.~Giani, M.~Pernici and P.~van Nieuwenhuizen,
``Gauged ${\cal N}=4$ $D = 6$ Supergravity,''
Phys.\ Rev.\ D {\bf 30}, 1680 (1984).

\bibitem{Cowdall:1998fn}
P.~M.~Cowdall,
``Supersymmetric electrovacs in gauged supergravities,''
Class.\ Quant.\ Grav.\  {\bf 15}, 2937 (1998)
[hep-th/9710214].

\bibitem{Kinar:2000vq}
Y.~Kinar, E.~Schreiber and J.~Sonnenschein,
``Q anti-Q potential from strings in curved spacetime: Classical results,''
Nucl.\ Phys.\ B {\bf 566}, 103 (2000)
[hep-th/9811192].

\bibitem{Sonnenschein:1999if}
J.~Sonnenschein,
``What does the string / gauge correspondence teach us about Wilson  loops?,''
hep-th/0003032.

\bibitem{Chamseddine:1997nm}
A.~H.~Chamseddine and M.~S.~Volkov,
``Non-Abelian BPS monopoles in ${\cal N} = 4$ gauged supergravity,''
Phys.\ Rev.\ Lett.\ {\bf 79} (1997) 3343
[hep-th/9707176].

\bibitem{Chamseddine:1998mc}
A.~H.~Chamseddine and M.~S.~Volkov,
``Non-Abelian solitons in ${\cal N} = 4$ gauged supergravity and 
leading order  string theory,''
Phys.\ Rev.\ D {\bf 57} (1998) 6242
[hep-th/9711181].


\bibitem{Bergman:1999na}
O.~Bergman, A.~Hanany, A.~Karch and B.~Kol,
``Branes and supersymmetry breaking in 3D gauge theories,''
JHEP {\bf 9910}, 036 (1999)
[hep-th/9908075], and references therein.

\bibitem{Dorey:2000rb}
N.~Dorey and D.~Tong,
``Mirror symmetry and toric geometry in three dimensional gauge theories,''
JHEP {\bf 0005}, 018 (2000)
[hep-th/9911094], and references therein.

\bibitem{Aharony:1997}
O.~Aharony, A.~Hanany, K.~Intriligator, N.~Seiberg and M.J.~Strassler,
``Aspects of N=2 Supersymmetric Gauge Theories in Three Dimensions,''
Nucl.\ Phys.\ B {\bf 499} (19997) 67 [hep-th/9703110].




\end{thebibliography}
\end{document}